\documentclass[%
 reprint,
 amsmath,amssymb,
 aps,
]{revtex4-2}

\usepackage{graphicx}
\usepackage{dcolumn}
\usepackage{bm}
\usepackage{color}
\usepackage{mathrsfs}
\usepackage{comment}
\usepackage{mathtools}
\usepackage{eurosym}
\usepackage{amsmath,amssymb}
\usepackage{graphicx}
\usepackage{dcolumn}
\usepackage{bm}
\usepackage{color}
\usepackage{multirow}
\usepackage{threeparttable}
\usepackage{adjustbox}

\usepackage{float}
\usepackage[english]{babel}

\usepackage{array}
\usepackage{dcolumn}
\usepackage{booktabs}
\usepackage[detect-all]{siunitx}
\usepackage{makecell}
\usepackage{xcolor}
\usepackage{yfonts}
\newcolumntype{P}[1]{>{\centering\arraybackslash}p{#1}}

\def\bc{\begin{center}}
\def\ec{\end{center}}

\def\beq{\begin{equation}}
\def\eeq{\end{equation}}

\begin{document}

\title{On superfluidity of indirect excitons in transition metals
trichalcogenides van der Waals heterostructures }
\author{ Roman Ya. Kezerashvili and Anastasia Spiridonova }
\affiliation{\mbox{$^{1}$Physics Department, New York
City College
of Technology, The City University of New York,} \\
Brooklyn, New York 11201, USA \\
\mbox{$^{2}$The Graduate School and University Center, The
City University of New York,} \\
New York, New York 10016, USA }
\date{\today}

\begin{abstract}
{We predict angle-dependent superfluidity for a new class of two-dimensional materials -- transition metal
trichalcogenides (TMTC). Within a mean-field approach superfluidity of indirect excitons in TMTC van der Waals heterostructures (vdWHs) is studied. We use different potentials for charged carriers interaction to analyze the
influence of the screening on the studied phenomena. Our study demonstrates
the angle-dependent superfluidity temperature in TMTC vdWHs: for a given density a maximum and minimum $T_c$ of superfluidity occurs along chain and $a$ directions, respectively. This work can guide
experimental research toward the realization of anisotropic superfluidity in TMTC
vdWH. We suggest an experiment for the observation of anisotropic superfluidity in TMTC vdWH. }
\end{abstract}

\maketitle


\section{Introduction}
Pairing of electrons and holes bound by the electrostatic attraction in a
system consisting of two spatially separated electron and hole layers was
proposed as a possible origin of superfluidity initially in coupled
semiconductor quantum wells \cite{Lozovik}. Following this, a very
substantial fraction of theoretical and experimental works have been devoted
to the study of Bose-Einstein condensation (BEC) and superfluidity of
indirect interlayer excitons (see reviews \citep{Snoke2013,Combescot}). After the discovery of graphene, electron-hole
superfluidity has been studied in different new class of two-dimensional\
(2D) materials \citep
{BKZGraf2012,LozovikGraf2012,Pernali2013,Zarenia2014,Fogler,MacDonald,Zarenia2016,BK2016,BK2017,Conti2019,Donck2020,Conti2021,Berm2022}.
On one hand, gapped graphene, Xenes, and TMDCs are 2D materials with isotropic
carrier particle masses, where superfluidity of a dipolar Bose gas is isotropic \citep
{BKZGraf2012,LozovikGraf2012,Pernali2013,Zarenia2014,Zarenia2016,Fogler,BK2016,BK2017,Donck2020,Conti2021}. On the other hand, phosphorene, group-V TMDCs (ReS$_2$ and ReSe$_2$), and TMTCs have asymmetric electron-hole masses along $x$ and $y$ directions
\citep{Ticknor2011,BGK2017,Zarenia2018,Zhang2019}. We cite these works, but the recent literature on the subject is not limited by them.

The anisotropic superfluidity for quasiparticles was first studied for He$%
^{3}$ \cite{Saslow}. The superfluid character of a dipolar BEC in a quasi-two-dimensional geometry and anisotropic superfluidity
in a superfluid system with anisotropic effective masses along principal three-dimensional (3D)
directions were investigated in Refs. \cite{Ticknor2011,Zhang2019}. There
has been a particular interest in the study of superfluidity of indirect
excitons in anisotropic van der Waals heterostructures (vdWH) with phosphorene sheets separated by thin hBN insulating layers. It
was predicted in Ref. \cite{BGK2017} the angular dependence of the mean-field critical temperature for superfluidity in a weakly interacting gas of dipolar excitons in phosphorene vdWH.
The anisotropic superfluidity is
predicted in Ref. \cite{Zarenia2018}, where it was found that the maximum
superfluid gap in the BEC regime is along the armchair direction. While a
special feature of phosphorene is the high in-plane anisotropy of its energy
band structure that leads to dipole excitons anisotropic superfluidity, phosphorene is unstable in air and highly toxic.

A stream of new 2D layered materials have been
developed over the past 10 years. Among them, there exist a small number of materials with strong in-plane
structural anisotropy such as transition metal trichalcogenides (TMTC): TiS$%
_{3}$, TiSe$_{3}$, ZrS$_{3}$, and ZrSe$_{3}$. In contrast to TiS$_{3}$, TiSe$%
_{3}$, ZrS$_{3}$, and ZrSe$_{3}$ monolayers are all indirect gap
semiconductors \cite{Jin2015,Ming2015}. There are several reasons to seriously
consider the TMTC family: they are environmentally friendly, have low
manufacturing cost of semiconductors, and forbid native oxide formation. These motivate us to study superfluidity in TMTC.

Heterostructures with two parallel TMTC monolayers with one $n$-doped and
the other a $p$-doped, respectively, separated by an insulating barrier of
stocked hBN monolayers to block electron-hole recombination \cite{hBN} is an
interesting platform for investigation of electron-hole superfluidity. It is assumed that 2D layers are populated with carriers of opposite polarity and equal density. In such
kind of platform due to electrostatic interaction electrons and holes can pair
to produce dipole (indirect) excitons, which are composite bosons~\cite%
{Comberscot}. This kind of the system was used to study the effect of superfluidity in
heterostructures composed with graphene, TMDC, and phosphorene monolayers
\cite{Fogler,MacDonald,BK2016,BK2017,BGK2017,Zarenia2018}.

In this paper, we predict superfluidity of indirect excitons in van der Waals heterostructures composed of TMTCs monolayers and demonstrate that the mean-field critical temperature is angle dependent: for a given density a maximum temperature of superfluidity occurs along chain direction while a minimum temperature occurs along the $a$ direction. In addition, we propose an experiment for observation of this phenomenon in TMTC heterostructure.

\section{Model, results and discussions}
To explore a feasibility of superfluidity of paired spatially separated
electrons and holes first consider the formation of indirect excitons (IXs), in
TMTC heterostructure. In the framework of the effective mass approximation, after the separation of the center of mass,
the Schr\"{o}dinger equation for the relative motion of an interacting
electron-hole pair with anisotropic masses in TMTC vdWH reads as
$\left[ -\frac{1}{2\mu _{x}}\frac{\partial ^{2}}{\partial x^{2}}-\frac{1}{%
2\mu _{y}}\frac{\partial ^{2}}{\partial y^{2}}+V(x,y)\right] \Phi
(x,y)=E\Phi (x,y),$
where $\mu _{x}=\frac{m_{x}^{e}m_{x}^{h}}{m_{x}^{e}+m_{x}^{h}}$ and $\mu
_{y}=\frac{m_{y}^{e}m_{y}^{h}}{m_{y}^{e}+m_{y}^{h}}$ are the anisotropic
reduced masses in the $x$ and $y$ directions, respectively, $%
m_{x}^{e}$ $(m_{x}^{h})$ and $m_{y}^{e}$ $(m_{y}^{h})$ correspond to the
electron (hole) effective masses along the $x$ and $y$ directions and $\Phi (x,y)$ is the electron-hole pair relative motion
wave function. 
However, for now note that, according to \citep{Saeed2017,Donck2018,Wang2020zrs,Tripathi2021}, the $a$ direction ($\theta=0$) corresponds to the $x$ direction, and the $b$ direction ($\theta=\pi/2$) corresponds to the $y$ direction.

The interaction potential $V(x,y)$ between the
electron and hole which resides in two different TMTC monolayers is:
$V_{C}\left( x,y\right) =-\frac{ke^{2}}{\kappa \left( \sqrt{x^{2}+y^{2}+D^{2}}%
\right) }$
in the case of the Coulomb electrostatic attraction or
\begin{eqnarray}
V_{RK}(x,y)=&&-\frac{\pi ke^{2}}{2\kappa \rho _{0}}\biggr[ H_{0}\left( \frac{\sqrt{x^{2}+y^{2}+D^{2}}}{\rho _{0}}\right) \nonumber \\
 && -Y_{0}\left( \frac{\sqrt{x^{2}+y^{2}+D^{2}}}{\rho _{0}}\right) \biggr]
\label{eq:indkeld}
\end{eqnarray}
if the indirect exciton is formed via the Rytova-Keldysh (RK) potential \cite%
{Rytova1967,Keldysh1979}. These two equations
describe the interaction between the electron and hole that are located in
two parallel TMTC monolayers separated by a distance $D=h+Nl_{\text{hBN%
}}$, where $l_{\text{hBN}}=0.333$ nm is the thickness of the hBN layer, $N$ is the number of hBN layers, and $%
h$ is the TMTC monolayer thickness. In Eq.~(\ref{eq:indkeld}) $e$ is the charge of the electron, $%
\kappa =(\epsilon _{1}+\epsilon _{2})/2$ describes the surrounding
dielectric environment, $\epsilon _{1}$ and $\epsilon _{2}$ are the
dielectric constants below and above the monolayer, respectively, $H_{0}$
and $Y_{0}$ are the Struve and Bessel functions of the second kind,
respectively, and $\rho _{0}$ is the screening length. The potential (\ref%
{eq:indkeld}) has the same functional form as one the derived in Ref. \cite%
{Cudazzo2011}, where the macroscopic screening is quantified by the 2D
polarizability. Following ~\cite{Cudazzo2011,Berkelbach2013} the screening
length $\rho _{0}$ can be written as $\rho _{0}=2\pi \chi _{2D}/\kappa $,
where $\chi _{2D}$ is the 2D polarizability. 2D layer polarizability can be
computed by standard first-principles technique \cite{Cudazzo2011} or
considered as a phenomenological parameter. Below for indirect excitons, we report
calculations for both the Coulomb and RK (\ref{eq:indkeld}) potentials to explore the role of the screening.

Our goal is to engineer vdWH with a high exciton binding energy. By averaging the Schr\"{o}dinger equation, one can
obtain the expectation value for the bound state
energies:
$E=\left\langle -\frac{1}{2\mu _{x}}\frac{\partial ^{2}}{\partial x^{2}}%
\right\rangle +\left\langle -\frac{1}{2\mu _{y}}\frac{\partial ^{2}}{%
\partial y^{2}}\right\rangle +\left\langle V(x,y)\right\rangle$.
The terms in this expression could be viewed as the sum of the average
values of the operators of kinetic and potential energies in 2D space
obtained by using the corresponding eigenfunction $\Phi (x,y)$. 
From this expression 
it follows that large reduced
masses $\mu _{x}$ and $\mu _{y}$ give small contributions of the kinetic
energy terms that lead to larger binding energy.
We examined vdWHs composed of two the same and two different TMTCs monolayers. Based on our results for binding energies of IXs, we have selected the following heterostructures where excitons have the largest binding energies:
ZrSe$_{3}/$ZrS$_{3}$, TiS$_{3}/$ZrS$_{3}$, TiS$_{3}/$ZrSe$_{3}$, ZrS$_{3}/$%
ZrS$_{3}$, ZrSe$_{3}/$ZrSe$_{3}$. The charge carrier in the first layer
is electrons and in the second one is holes. The binding energies of excitons
increase from about 73 to 89 meV for the RK potential and from about 105 to 133 meV for the Coulomb potential when TMTC monolayers are separated
by a single hBN layer. The increase of the hBN layers reduces the binding
energies of excitons monotonically. For six hBN layers the energies vary
from 49 to 57 meV for the RK potential and from 58 to 68 meV for the Coulomb
potential. Thus, the difference between excitons binding energies obtained with the Coulomb and RK potentials decreases. The binding energies of indirect excitons in vdWHs are given in Table \ref{tab:energy_het} in Appendix A. Thus, we can
conclude that a dilute gas of
bound excitons can be induced in TMTC vdWHs.

We now turn our attention to a dilute gas of bound
electrons and holes in
TMTC vdWHs, when $nr_{X}^{2}\ll 1$, where $n$ and  $r_{X}$ are the concentration and radius for indirect excitons, respectively. Under this condition,
we treat the dilute system of indirect excitons in TMTC vdWH as a
weakly interacting Bose gas.

The excitons are 
composite bosons~\cite{Comberscot} and at large interlayer separations, the
exchange effects ~\cite{Moskalenko_Snoke,Snoke2013} in the exciton-exciton interactions in a TMTC vdWHs can be
neglected since the exchange interactions in vdWH are suppressed due to the low tunneling
probability, caused by the shielding of the dipole-dipole interaction by the
insulating barrier. When an electron and hole are located in
two different monolayers, the electron-hole forms an energetically favorable configuration
where dipoles are parallel to each other.

Consider excitons as composite bosons~\cite{Comberscot} forming a weakly interacting Bose
gas of direct excitons in TMTC vdWH. A general form of a many-body Hamiltonian
for the 2D interacting IXs in second quantization reads as
\cite{Lifshitz,Pitaevsky2016}
\begin{equation}
\hat{H}=\sum \varepsilon _{0}(p,\theta )a_{\mathbf{p}}^{\dagger }a_{\mathbf{p%
}}+\frac{g}{S}\sum a_{\mathbf{p}_{1}}^{\dagger }a_{\mathbf{p}_{2}}^{\dagger
}a_{\mathbf{p}_{1}}a_{\mathbf{p}_{2}},
\label{Ham}
\end{equation}%
where summation of all impulses appearing in indices, $a_{\mathbf{p}%
}^{\dagger }$ and $a_{\mathbf{p}}$ are Bose creation and annihilation
operators for IXs with momentum $\mathbf{p}$, $S$ is a
normalization area for the system, $g$ is a coupling constant for the
interaction between two IXs, and $\varepsilon _{0}(p,\theta )$
is the angle-dependent energy spectrum of non-interacting indirect excitons
$\varepsilon _{0}(\mathbf{p})=\frac{p_{x}^{2}}{2M_x}+\frac{p_{y}^{2}}{%
2M_y}=\frac{p^{2}}{2M_{0}(\theta )}$, where $p_{x}=p\cos \theta $ and $%
p_{y}=p\sin \theta $ are the polar coordinate for the center-of-mass momentum $\mathbf{p}$
and $M_{0}(\theta )=\left[ \frac{\cos ^{2}\theta }{M_x}+\frac{\sin
^{2}\theta }{M_y}\right] ^{-1}$ is angle-dependent effective mass, where $M_x=(m_x^e+m_x^h)$ and $M_y=(m_y^e+m_y^h)$. The only
difference between (\ref{Ham}) and the Hamiltonian for isotropic
excitons of weakly interacting Bose gas is that the single-particle energy
spectrum of non-interacting excitons is angle-dependent due to the angular
dependence of the exciton effective mass $M_{0}(\theta )$.

Consider a weakly interacting gas of IXs in a TMTC heterostructure in the framework of the Bogoliubov approximation. The
chemical potential of the interacting dilute Bose gas is given by $\mu =gn$
\cite{Abrikosov,Lifshitz,Pitaevsky2016} and following the standard text book
procedure \cite{Pitaevsky2016}, one obtains
\begin{equation}
 \hat{H}=E_{0}+\sum\limits_{%
\mathbf{p\neq 0}}\epsilon (p,\theta )b_{\mathbf{p}}^{\dagger }b_{\mathbf{p}},
\end{equation}
where $E_{0}$ is the ground state energy \cite{Pitaevsky2016}, $%
b_{p}^{\dagger }$ and $b_{p}$ are the new set of creation and annihilation
Bose operators for the quasiparticles that follow the same Bose commutation
relations $b_{\mathbf{p}}b_{\mathbf{p}^{^{\prime }}}^{\dagger }-$ $b_{%
\mathbf{p}^{^{\prime }}}^{\dagger }b_{\mathbf{p}}=\delta _{\mathbf{pp}%
^{^{\prime }}}$ obeyed by the original particle operators $a_{\mathbf{p}%
}^{\dagger }$ and $a_{\mathbf{p}},$
and
\begin{eqnarray}
\epsilon (p,\theta )=&&\left[ \frac{gn}{M_{0}(\theta )}p^{2}+\left( \frac{p^{2}%
}{2M_{0}(\theta )}\right) ^{2}\right] ^{1/2} \nonumber \\
=&&\left[ \left( \varepsilon
_{0}(p,\theta )+\mu \right) ^{2}-\mu ^{2}\right] ^{1/2}  \label{Bog}
\end{eqnarray}%
is the famous Bogoliubov dispersion law for the elementary excitations of
the system obtained in 1947 \cite{Bogoliubov}. In our case, due to the mass anisotropy, the spectrum of collective excitations $\varepsilon _{0}\left( p,\theta \right) $ depends on
the angle $\theta $, while
the exciton-exciton interaction term is angle-independent.
Thus, the angle-dependent spectrum of the collective excitations $\epsilon
(p,\theta )$ turns out to be uniquely fixed by the interaction between two
dipolar excitons, i.e. by the coupling constant $g$. The derivation of the coupling constant $g$ is given in Appendix B.

Introduce the mean-field critical temperature $T_c(\theta)$. At nonzero temperatures, the density of the superfluid component $\rho
_{s}(T)$ is defined as $\rho _{s}(T)=\rho -\rho _{n}(T)$, where $\rho $ is
the total 2D density of the exciton gas and $\rho _{n}(T)$ is the density of the
normal component \cite{Pitaevskii}. Suppose that the excitonic system moves
with a velocity $\mathbf{u}$, which means that the superfluid component
moves with the velocity $\mathbf{u}$. To calculate the density of the superfluid component
consider the total density flow $\mathbf{J}$ \cite{Bardeen}
for a Bose gas of quasiparticles. Following Ref. \cite{Saslow} and restricting oneself to
the first order term, in the reference frame where the superfluid
component is at rest, one obtains
$\mathbf{J}=-\frac{s}{k_{B}T}\int \frac{d^{2}p}{(2\pi \hbar )^{2}}\mathbf{p}%
\left( \mathbf{pu}\right) \frac{\partial f\left[ \varepsilon (p,\theta )%
\right] }{\partial \varepsilon (p,\theta )}$
. In the latter expression $f\left[ \varepsilon (p,\theta )\right] =\left( \exp %
\left[ \varepsilon (p,\theta )/(k_{B}T)\right] -1\right) ^{-1}$ is the
Bose-Einstein distribution function for quasiparticles with anisotropic
quasiparticle energy $\varepsilon (p,\theta )$, $s=4$ is the spin degeneracy
factor, and $k_{B}$ is the Boltzmann constant. The normal density $\rho _{n}$
for the anisotropic system has a tensor form \cite{Saslow} and is defined through the component of density flow $\mathbf{J}$ vector as
$J_{i}=\rho_{n}^{ij}(T)u_{j}=\frac{n_{n}^{ij}(T)}{M_{j}}u_{j}$,
where $i$ and $j$ denote either the $x$ or $y$ components of the vector, and $n_{n}^{(ij)}(T)$ are the tensor elements of normal concentration. In
a superfluid with an anisotropic quasiparticle energy, the normal-fluid
concentration is a diagonal tensor \cite{Saslow} such that $n_{n}^{xy}(T) =n_{n}^{yx}(T)=0$, and expressions for $n_{n}^{xx}$ and $n_{n}^{yy}$ are given in Appendix C.
At last, the concentration of the normal component is%
\begin{equation}
n_{n}(\theta ,T)=\sqrt{n_{n}^{xx}(T)^{2}\cos ^{2}\theta
+n_{n}^{yy}(T)^{2}\sin ^{2}\theta }.  \label{nNormal}
\end{equation}%
Finally, the mean-field critical temperature $T_{c}(\theta )$ of the phase transition
related to the occurrence of superfluidity 
is determined by the condition $n_{n}(\theta,T_{c}(\theta ))=n$, where $n$ is a total exciton concentration.

\begin{figure*}[htp]
\includegraphics[width=170mm]{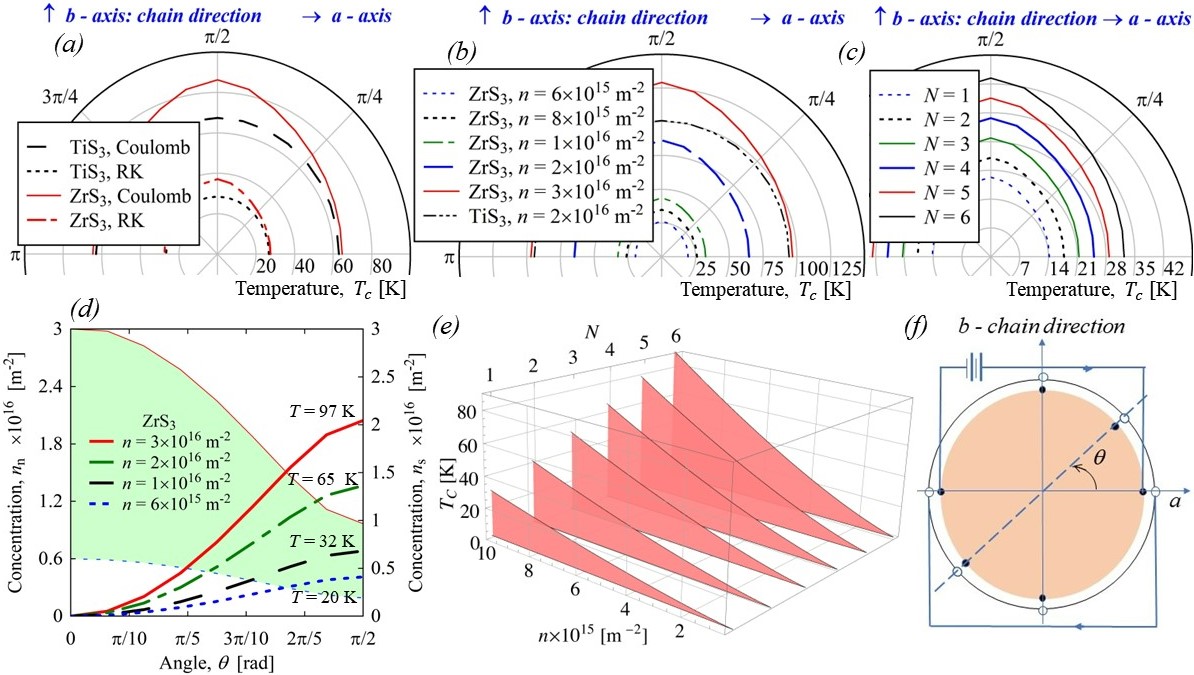}
\caption{The contour plot for the angle-dependent mean-field critical
temperature for superfluidity $T_{c}$ in polar coordinates for IXs (a) when $n=3$ $\times$ $10^{16}$ and $n=8$ $\times$ $10^{15}$ m$^{-2}$ in ZrS$_{3}$ vdWH (solid and dotted-dashed curves, respectively) and TiS$_{3}$ vdWH (dashed and dotted curves, respectively) using the Coulomb and RK potentials and $N=6$; (b) $T_{c}$ at different excitonic concentrations in ZrS$_{3}$ vdWH. The
calculation for TiS$_{3}$ vdWH is performed at $n=2$ $\times$ $10^{16}$ m$^{-2}$ using the Coulomb potential and $N = 6$. (c) $T_c$ for different numbers of hBN layers when $%
n=1$ $\times$ $10^{16}$ m$^{-2}$. The calculations are performed using the
Coulomb potential. (d) Thick curves show the angle dependence of the superfluid concentration for different total concentrations of the excitonic gas at the corresponding $T_{c}$ at which the superfluidity vanishes along the $a$ axis. Two thin curves bound green shaded region that shows the angle-dependence of the normal concentration. (e) $T_{c}$
for indirect excitons in TiS$_{3}$ vdWHs 
in the limit of the sound-like spectrum of collective excitations
as a function of concentration and interlayer separation. The calculations are
performed for the Coulomb potential. (f) A schematic plan-view for a suggested experiment. The "active" and parallel "passive" layers are shown as shaded and white areas, respectively. Filled and open dots indicate contacts to the "active" and "passive" layers, respectively. }
\label{fig:AngDep}
\end{figure*}

Our main results are summarized in the contour plots presented in Fig. \ref%
{fig:AngDep}(a)-\ref{fig:AngDep}(c), where we report $T_{c}$
for superfluidity with the spectrum of collective excitations given by Eq. (\ref{Bog}). The results for sound like spectrum are given in Appendix D. While 
examining $T_{c}$
obtained with the Bogoliubov energy spectrum (\ref{Bog}) and the sound like spectrum (Appendix C), we have found that vdWH that are composed of the same TMTCs monolayers have higher $T_c$ then vdWHs composed from two different TMTCs monolayers.

In calculations, we consider the excitonic concentrations in the range of 6 $\times$ $10^{15}$ to 3 $\times$ $10^{16}$ m$^{-2}$. The indirect excitons concentrations 6 $\times$ $10^{15}$ m$^{-2}$ are achievable in 2D TMDC heterostructures. For instance, the estimated IXs concentration for the optimal indirect excitons propagation condition in Ref. \cite{Fowler2022} is $n$ $\sim$ 6 $\times$ $10^{15}$ m$^{-2}$. The concentrations $\sim 2-3$ $\times 10^{16}$ m$^{-2}$ can be achieved for indirect excitons in 2D TMDC heterostructures, and order of $1-8$ $\times10^{16}$ m$^{-2}$ in single-layer black phosphorus \cite{Surrente2016}. However, one should mention that high densities approach the Mott transition \cite{Fogler}, beyond which the hydrogen-like indirect excitons vanish and the Cooper-pair like excitons may form. The latter were observed in GaAs heterostructure \cite{Choksy2022}. In Fig. \ref{fig:AngDep}(a) results of
calculations for $T_{c}$ using the Coulomb and RK potentials are presented for ZrS$_{3}$ and TiS$_{3}$ vdWHs. $T_{c}$ obtained using the Coulomb potential is more than twice as big as one calculated with the RK potential. This difference is due to the strong screening of the RK potential. Interestingly enough, the ratios of $T_{c}$ along the chain direction ($b$ axis) to $T_{c}$ along the $a$ axis do not depend on the potential and are 1.07 and 1.33 for TiS$_{3}$ and ZrS$_{3}$ vdWH, respectively. Thus, ZrS$_{3}$ demonstrates a stronger angle-dependent $T_{c}$. It is interesting to compare T$_c$ for ZrS$_3$ and phosphorene. For example, in the case of phosphorene T$_c$ at the density $3$ $\times10^{16}$ has maximum at $\theta=0$ and minimum when $\theta=\pi/2$ \cite{BGK2017}, in contrast to ZrS$_3$ where T$_c$ has minimum at $\theta=0$ and maximum at $\theta=\pi/2$. 

In Fig. \ref{fig:AngDep}(b) results of
calculations are presented for ZrS$_{3}$ vdWH for different concentrations of
exciton gas and one data set for TiS$_{3}$ vdWH when $n=2$ $\times$ $10^{16}$ m$^{-2}$. From the polar plot, it can clearly be
seen that for ZrS$_{3}$ the critical temperature of superfluidity is
aligned along the chain direction ($b$ axis) for all considered excitonic
concentrations. $T_{c}$ increases as the concentration increases. The maximum $T_{c}$
is observed along the chain direction at $\theta =\pi /2$,
while minimum\ $T_{c}$ is observed along the $a$ axis ($\theta=0$). An interesting observation is that the
ratio $T_{c\text{ max}}/T_{c\text{ min}}$ almost does not depend on the excitonic concentration
and decreases monotonically from 1.34 at $n=6\times 10^{15}$ m$^{-2}$ to
1.33 at $n=3\times 10^{16}$ m$^{-2}$. The behavior of $T_{c}$
for TiS$_{3}$ appears almost isotropic with $T_{c\text{ max}}/T_{c\text{ min}} = 1.07$. This phenomenon can be explained by the fact that $\mu_x$ and $\mu_y$ are almost equal to each
other (very small mass anisotropy 
along $x$ and $y$).
In addition, as can be seen in Fig. \ref{fig:AngDep}(c), the increase in the number of
insulating hBN layers, $N$, leads to the increase of $T_{c}$.

As stated above, our studies reveal that the maximum $T_{c}$ occurs along the chain direction. This can be explained by TMTCs monolayer structure. The simplest description of geometrical structures known as
quasi-one-dimensional compounds is usually in terms of the condensation of
small clusters of atoms into infinite chains. In TMTCs, trigonal prismatic
units MX$_{6}$ with six atoms of chalcogenides and one atom of transition
metal condense laterally into the two-dimensional layers of MX$_{3}$ and
longitudinally into the one-dimensional linear chains of the MX$_{6}$. In other words, these compounds crystallize in a
chain structure \cite{Hodeau78} composed of wedge-shaped trigonal prisms
stacked one on top of the other, base to base, while the metal atom at the
center of each prism is coordinated to six chalcogen atoms at the corners of
the prisms.

For instance, ZrS$_{3}$ has a rather strong in plane
structural anisotropy due to weak chain-chain interaction. Each ZrS$_{3}$ layer consists of 1D chains made of MX$_{6}$ octahedrons extending along the $b$ direction \citep{Ait94,Pant2016, Jin2015} that are weakly coupled laterally in the $a$ axis direction where the $M$-$X$ linkages between chains via two chalcogen atoms in neighboring chains are held
together weakly by forces of the van der Waals type. Inside the 1D prismatic chain, the unit orientated along $b$, six atoms of S are equally
separated from the metal atom Zr at $\sim 2.63$ \AA , but the S atoms
outside of the chain are slightly farther away $\sim 2.75$ {\AA} \cite{Pant2016}. This difference in bonds' length
results in slightly decoupled 1D-chain like structures confined in the $ab-$%
plane. It is worth noting the results for bonds' length stated above are consistent with earlier
structural studies on ZrS$_{3}$ \cite{Wieting79,Wieting80, Jin2015}. The conclusion drawn
from the study \cite{Wieting79} was that the strength of the coupling
between the chains was approximately one-fifth of the strength of the M$-$X
bond within the chain. Thus,
our results indicate that the maximum $T_c$ 
occurs along the quasi-one-dimensional linear chains of the ZrS$%
_{3}$.

In Fig. \ref{fig:AngDep}(d) using thick lines, we show the dependence of the superfluid concentration, $n_s$, on the angle for different total concentrations, $n$, of the dipole excitons at temperatures at which the superfluidity is killed. In other words, we take $T_{c}$ at which superfluidity is killed when $\theta = 0$ ($a$ axis), then we keep $T_c$ constant and increase $\theta$ and calculate $n_s$. The concentration of the superfluid component increases with the increase of $\theta$ and achieves the maximum at $\theta = \pi/2$, which corresponds to chain direction. In addition, we show the angular dependence of the normal concentration $n_n$ that falls within a green shaded area between the thin red solid and dashed blue curves obtained for $n = 3$ $\times$ $10^{16}$ m$^{-2}$ at T$_c$=97 K and $n = 6$ $\times$ $10^{15}$ m$^{-2}$ at T$_c$=20 K, respectively.

In our calculations with RK potential, we use the isotropic approximation for the 2D polarizability. In general, 2D polarizability of anisotropic materials should be considered as a tensor. However, following Ref. \cite{Rodin2014}, we used rather crude approximation $\chi_{{2D}_{xx}}=\chi_{{2D}_{yy}}$ and replaced both by their average. In Refs. \citep{Torun2018,Donck2018,KezSpir2022} the isotropic approximation of 2D polarizability was also used. This approximation simplifies the calculation in case of RK potential and did not affect substantially exciton binding energy \cite{Rodin2014}, in contrast of electron-hole mass anisotropy.

In the limit of small momenta $p$, when $\varepsilon _{0}(p,\theta )\ll gn$,
we expand the spectrum of collective excitations $\epsilon (p,\theta )$
up to first order with respect to the momentum $p$ and obtain the sound mode
of the collective excitations $\epsilon (P,\theta )=c_{s}(\theta )P$, where $c_{s}(\theta )$ is the angle-dependent sound velocity given in Appendix D. Since at low momenta the sound-like energy spectrum of collective excitations of the IXs in a TMTC heterostructure satisfies the Landau criterion for superfluidity, the superfluidity of IXs in this system is possible. Thus, such systems below the Berezinskii-Kosterlitz-Thouless (BKT) transition temperature may support a superflow
of indirect excitons. One can estimate the BKT phase transition temperature from the condition  $n_{n}(T)=n$. The latter leads to
\begin{equation}
T_{BKT}=\left( \frac{\pi (\hbar g)^{2}}{2\zeta (3)s\sqrt{M_{x}M_{y}}}\right)
^{1/3}\frac{n}{k_{B}}.  \label{Tcs}
\end{equation}
and can be considered as an upper bound to the superfluid transition temperature.

Exclusivity of TiS$_{3}$ monolayer which has a direct band gap motivates us to calculate $T_{c}$ for TiS$_{3}$ vdWH in the limit for the sound like spectrum of collective excitations.
The 
$T_{c}$ for IXs in
TiS$_{3}$ vdWH as a function of concentration and number of hBN insulating
layers is shown Fig. \ref{fig:AngDep}(e). Calculations are performed with the Coulomb potential. $T_{c}$ increases with the increase of the
concentration and interlayer separation.

The angle-dependent superfluidity in TMTC vdWHs may be observed in electron-hole Coulomb drag experiments \cite{DragC}. 
In such experiment, a voltage is applied to the one of the layers, known as the "active" layer. The other "passive" layer is closed upon itself by connecting to an external resistor. The electric current in the active layer induces a current in the passive layer in the opposite direction by means of the Coulomb drag, i.e., by the momentum transfer due to the interlayer electron-hole interaction. The drag current is typically much smaller than the drive current owing to the heavy screening of the Coulomb interaction \cite{Nandi2012}.

A schematic plan view for the suggested experiment is shown in Fig. \ref{fig:AngDep}(f). The geometry of the experiment revealing the existence of the exciton angle-dependent superfluidity will include measurements at the angle $\theta = 0$ with respect to the $a$ axis at the corresponding transition temperature $T_{c}$ for the given excitonic concentration, $n$, when there is no superfluidity. At the next step, one applies voltage to "active" layer at the angle $\theta$ and make measurement in the "passive" layer at the same angle. By keeping $T_{c}$ constant and gradually increasing the angle $\theta $ with respect to the $a$ axis and making measurements of the induced current in the "passive" layer, let us say at $\theta = \pi/6$, $\theta = \pi/4$, $\theta = \pi/3$ and $\theta = \pi/2$ (at the same angles the voltage is applied to the "active" layer), one can observe the increment of the induced current with the increase of $\theta$. The maximum value is achieved in the chain direction 
at $\theta = \pi/2$. In such measurements, due to the electron-hole mass anisotropy, there is no induced current at $T_{c}$, which corresponds to the $a$ axis direction, and there is the maximal current at this temperature in the chain direction. If instead of the superfluid current, the normal current is measured at $T_{c}$ when $n_{c}$ has the minimum based on results in Fig. \ref{fig:AngDep}(f) one will observe the opposite picture: the normal current depletion from the maximum at $\theta=\pi/2$ to the minimum at $\theta = 0$.
While the purely Coulomb drag can be the most qualitative features of the effect, other mechanisms of momentum transfer can also contribute to the observed behavior \cite{DragC}.

\section{Conclusions}
To conclude, in the framework of a mean-field approach within the Bogoliubov
approximation, we predict angle-dependent superfluidity of indirect excitons in van der Waals heterostructures composed from a new class of two-dimensional materials: TMTC. The angle-dependent superfluidity is due to the directional anisotropy of the electron and hole effective masses. The angle dependence of $T_{c}$ occurs beyond the sound-like approximation for the spectrum of collective excitation: for a given density a maximum and minimum $T_c$ of superfluidity is along the chain and $a$ directions, respectively. Thus, in contrast to the anisotropic behavior of $T_{c}$ for  phosphorene \cite{BGK2017}, the vastly different angular dependence of $T_{c}$ is observed for ZrS$_{3}$.
In calculations, we used the Rytova-Keldysh and Coulomb potentials for charged carriers interaction to analyze the
influence of the screening. For both potentials the angle-dependent superfluidity of indirect excitons is observed. The critical temperature for the phase transition obtained using the Coulomb potential is significantly larger than $T_{c}$ calculated with the RK potential. Thus, our results demonstrate that the screening does significant affect the phase transition temperature. We have reported binding energies for indirect excitons in vdWH composed of two different and the same TMTC monolayers. The indirect excitons bound strongly enough and can induce a dilute excitonic gas. However, we have found that heterostructures composed of the same TMTCs monolayers have higher $T_c$ then vdWH composed of two different TMTC monolayers. Also, $T_{c}$ of superfluidity in TMTC vdWHs can be manipulated by varying the excitonic density and the number of hBN insulated layers. Finally, we suggest and discuss the possibility of the experimental observation of the angle-dependent superfluidity
via Coulomb drag experiments.

Our study demonstrates the angle-dependent superfluidity in TMTC heterostructures, and we hope that it will motivate future experimental and theoretical investigations on excitonic BEC and anisotropic superfluidity in TMTC heterostructures.

\textbf{Acknowledgments.} The authors are thankful to L. V. Butov for the useful discussion. This work is supported by the U.S. Department of Defense under Grant No. W911NF1810433.

\appendix
\setcounter{figure}{0} \renewcommand{\thefigure}{A.\arabic{figure}}

\section{Binding energies of indirect excitons in TMTC vdWH} \label{sup_mat:eneriges}

We calculated the binding energies of indirect excitons in TMTC vdWH using the Coulomb and Rytova-Keldysh potentials. TMTC van der Waals heterostructures that are considered consist from either from the same two TMTC monolayers or two different TMTC monolayers separated by hBN monolayers. Results for the binding energies of indirect excitons in ZrSe$_{3}/$ZrS$_{3}$, TiS$_{3}/$ZrS$_{3}$, TiS$_{3}/$ZrSe$_{3}$, ZrS$_{3}/$%
ZrS$_{3}$, ZrS$_{3}/$%
ZrSe$_{3}$, ZrSe$_{3}/$ZrSe$_{3}$ vdWH are presented in Table \ref{tab:energy_het}.
\begin{table*}[htp]
\caption{Binding energies of indirect excitons in TMTC van der Waals
heterostructures. Energies are given in meV and electron and hole masses in $x$ and $y$ directions are given in units of the free electron mass}
\label{tab:energy_het}
\centering
\begin{adjustbox}{width=1\textwidth}
 \begin{threeparttable}[htp]
\begin{tabular}{c|cc|cc|cc|cc|cc|cc}
\hline\hline
& Electron & Hole & Electron & Hole   & Electron & Hole   & Electron & Hole & Electron & Hole & Electron & Hole \\ \hline
& ZrSe$_3$\tnote{a} & ZrS$_3  $\tnote{a}  & TiS$_{3}$\tnote{b}  & ZrS$_{3}   $\tnote{a} & TiS$_3$\tnote{b}  & ZrSe$_3  $\tnote{a }  & ZrS$_{3}$\tnote{a}  & ZrS$_{3}$\tnote{a}  & ZrS$_{3}$\tnote{a}  & ZrSe$_{3}   $\tnote{a  } & ZrSe$_{3}$\tnote{a}  & ZrSe$_{3}   $\tnote{a  } \\ \hline
$m_x$ & 0.16 & 1.28 & 1.52 & 1.28 & 1.52 & 2.36 & 1.3 & 1.28 & 1.3 & 2.36 & 0.16 & 2.36\\
$m_y$ & 6.72 & 0.42 & 0.4 & 0.42 & 0.4 & 0.89 & 0.4 & 0.42 & 0.4 & 0.89 & 6.72 & 0.89 \\ \hline\hline
Potential & $V_{RK}$ & $V_C$ & $V_{RK}$ & $V_C$ & $V_{RK}$ & $V_C$ & $V_{RK}$
& $V_C$ & $V_{RK}$ & $V_C$ & $V_{RK}$ & $V_C$ \\ \hline
1 & 72.8 & 105.3 & 87.7 & 123.2 & 87.5 & 132.8 & 88.6 & 121.8 & 87.7 & 130.0 & 73.7 & 114.5\\
2 & 66.5 & 89.9  & 78.9 & 103.6 & 79.0 & 110.9 & 79.6 & 102.6 & 79.2 & 109.0 & 67.5 & 97.1 \\
3 & 61.2 & 78.7  & 71.7 & 89.7 & 72.0  & 95.6  & 72.2 & 88.9  & 72.1 & 94.1 & 62.2 & 84.6 \\
4 & 56.7 & 70.2  & 65.7 & 79.2 & 66.1  & 84.2  & 66.0 & 78.6  & 66.2 & 83.0 & 57.7 & 75.1 \\
5 & 52.7 & 63.5  & 60.6 & 71.1 & 61.2  & 75.3  & 60.9 & 70.6  & 61.2 & 74.3 & 53.8 & 67.7 \\
6 & 49.3 & 58.0  & 56.2 & 64.6 & 56.9  & 68.2  & 56.4 & 64.1  & 56.9 & 67.4 & 50.3 & 61.6 \\ \hline\hline
\end{tabular}%

\begin{tablenotes} \footnotesize
            \item [a] Reference \cite{Jin2015}
            \item [b] Reference \cite{Donck2018}
        \end{tablenotes}
\end{threeparttable}
\end{adjustbox}
\end{table*}

\section{The coupling constant $g$}
The interaction parameters for the exciton-exciton coupling in very
dilute systems could be obtained assuming the exciton-exciton repulsion exists only at distances between excitons greater than the distance
from the exciton to the classical turning point \cite{BKKL}. Following Ref.\cite{BKKL}, one can
obtain the coupling constant for the exciton-exciton interaction:
\begin{equation}
g=\int_{R_{0}}^{\infty }V_{\text{dd}}\left( R\right) RdRd\theta ,
\end{equation}
where $V_{%
\text{dd}}\left( R\right) $ is the dipole-dipole interaction and $R_{0}$
corresponds to the distance between two IXs at their classical
turning point. $R_{0}$, is determined by the
conditions that the energy of two excitons cannot exceed
the doubled chemical potential $\mu $ of the system~\cite{BKKL}, i.e., $V_{%
\text{dd}}(R_{0})\approx 2\mu.$ The last expression is
reasonable for a weakly-interacting Bose gas of IXs.

To find  $V_{%
\text{dd}}(R)$, consider dipole-dipole interaction as the interactions of
positive and negative charges of one dipole with another dipole. Based on
well known screened 2D potential, the Rytova-Keldysh potential, $V_{RK}$, \citep{Rytova1967, Keldysh1979}, which is widely used for the description of excitonic complexes in 2D materials \cite{Kezerashvili}, a dipole-dipole
interaction is obtained in Ref. \cite{RK_VK2022}. The explicit analytical
form for the dipole-dipole interaction in 2D configuration space for two excitons has the following form \cite{RK_VK2022}:

\begin{equation}
\begin{split}
V_{\text{dd}}\left( R\right) &=-\frac{\pi k}{2\kappa \rho _{0}}\biggl\{ \left[
H_{-1}\left( \frac{R}{\rho _{0}}\right) -Y_{-1}\left( \frac{R}{\rho _{0}}%
\right) \right] \frac{\mathbf{\mathbf{d}_{1}\mathbf{\cdot }d_{2}}}{\rho _{0}R%
} \\
& + \left[ H_{-2}\left( \frac{R}{\rho _{0}}\right) -Y_{-2}\left( \frac{R}{\rho
_{0}}\right) \right] \frac{\mathbf{R\mathbf{\cdot }d_{1}R\cdot d}_{2}}{\rho
_{0}^{2}R^{2}}\biggl\} ,  \label{DDK}
\end{split} 
\end{equation}
where $H_{-1}$, $H_{-2}$ and $Y_{-1}$, $Y_{-2}$ are the Struve functions and Bessel functions of the second kind, respectively. For dipolar excitons formed between two parallel layers separated by distance $D$ the latter expression  becomes: 
\begin{equation} \label{eq:vdd_rk}
V^{RK}_{\text{dd}}\left( R\right) =-\frac{\pi k}{2\kappa \rho _{0}}\left[
H_{-1}\left( \frac{R}{\rho _{0}}\right) -Y_{-1}\left( \frac{R}{\rho _{0}}%
\right) \right] \frac{d^{2}}{\rho _{0}R},
\end{equation}
where $d=eD$ is a dipole moment of the IX. For the Coulomb potential
the dipole-dipole interaction of IXs are well known:
\begin{equation}\label{eq:vdd_c}
V_{\text{dd}}^{C}\left( \mathbf{R}\right) =\frac{k}{\kappa }\frac{d^{2}}{%
R^{3}}.
\end{equation}
After that, it is easy to obtain $R_{0}$ for $V_C$:
\begin{equation} \label{eq:r_c}
R_{0}=\frac{1}{2\sqrt{\pi n}}
\end{equation}
For the RK potential, we obtain the following equation for $R_{0}$:
\begin{equation} \label{eq:r_rk}
4\pi n\rho _{0}^{2}y\left[ H_{0}(y)-Y_{0}(y)\right] =-\left[
H_{-1}(y)-Y_{-1}(y)\right],
\end{equation}
where $y=\frac{R_0}{\rho_0}$.

Using the expressions for $V_{\text{dd}}\left( R\right)$ [Eqs. (\ref{eq:vdd_rk}) and (\ref{eq:vdd_c})] and $R_0$ [Eqs. (\ref{eq:r_c}) and (\ref{eq:r_rk})], one obtains the exciton-exciton
coupling constant $g$:

\begin{equation}\label{DDK2}
g=
\begin{cases}
	\frac{\pi ^{2}ke^{2}D^{2}}{\epsilon _{d}\rho _{0}}\left[ H_{0}\left( \frac{R_{0}}{\rho _{0}}\right) -Y_{0}\left( \frac{R_{0}}{\rho_{0}}\right) \right], & \text{for} \quad V_{RK}(x,y), \\
	\frac{2\pi ke^{2}D^{2}}{\epsilon _{d}R_{0}}=\frac{4\pi ke^{2}D^{2}\sqrt{\pi n}}{\epsilon _{d}}, & \text{for} \quad V_C(x,y).
 \end{cases}
\end{equation}%

\section{The angle-dependent normal component concentration in an anisotropic system}\label{sup_mat:angle-dep}
The normal density $\rho _{n}$
for the anisotropic system has tensor form \cite{Saslow} and defined through the component of density flow $\mathbf{J}$ vector as
\begin{equation}
J_{i}=\frac{n_{n}^{ij}(T)}{M_{j}}u_{j}.  \label{Component}
\end{equation}%
where $i$ and $j$ denote either the $x$ or $y$ components of the vector and $n_{n}^{(ij)}(T)$ are the tensor elements of normal component concentration. In
a superfluid with an anisotropic quasiparticle energy, the normal-fluid
concentration is a diagonal tensor with $n_{n}^{xx}$ and $n_{n}^{yy}$ given
by \cite{Saslow}
\begin{widetext}
\begin{eqnarray}
n_{n}^{xx}(T) &=&\frac{s}{k_{B}T}\frac{1}{M_{x}}\frac{1}{(2\pi \hbar )^{2}}%
\int d\mathbf{p}\frac{\exp \left[ \varepsilon (p,\phi )/(k_{B}T)\right] }{%
\left( \exp \left[ \varepsilon (p,\phi )/(k_{B}T)\right] -1\right) ^{2}}%
\cos ^{2}\phi ,  \label{nxx} \\
n_{n}^{yy}(T) &=&\frac{s}{k_{B}T}\frac{1}{M_{y}}\frac{1}{(2\pi \hbar )^{2}}%
\int d\mathbf{p}\frac{\exp \left[ \varepsilon (p,\phi )/(k_{B}T)\right] }{%
\left( \exp \left[ \varepsilon (p,\phi )/(k_{B}T)\right] -1\right) ^{2}}%
\sin ^{2}\phi ,  \label{nyy} \\
n_{n}^{xy}(T) &=&n_{n}^{yx}(T)=0.  \label{3}
\end{eqnarray}%
Thus, the concentration of the normal component $n_{n}(\theta ,T)$ is%
\begin{equation}
n_{n}(\theta ,T)=\sqrt{n_{n}^{xx}(T)^{2}\cos ^{2}\theta
+n_{n}^{yy}(T)^{2}\sin ^{2}\theta }.  \label{nNormal2}
\end{equation}%
The mean-field critical temperature $T_{c}(\theta )$ of the phase transition
related to the occurrence of superfluidity 
is determined by the condition%
\begin{equation}
n_{n}(\theta ,T_{c}(\theta ))=n.  \label{Condition}
\end{equation}
\end{widetext}

\section{The sound mode of the collective excitations} \label{sup_mat:sound}
In the limit of small momenta $p$, when $\varepsilon _{0}(p,\theta )\ll gn$,
we expand the spectrum of collective excitations $\varepsilon (p,\theta )$
up to first order with respect to the momentum $p$ and obtain the sound mode
of the collective excitations $\varepsilon (p,\theta )=c_{s}(\theta )p$,
where $c_{s}(\theta )$ is the angle-dependent sound velocity, given by
\begin{equation}
c_{s}(\theta )=\sqrt{\frac{gn}{M_{0}(\theta )}}.
\end{equation}

\begin{figure}[h]
\includegraphics[width=5.3cm]{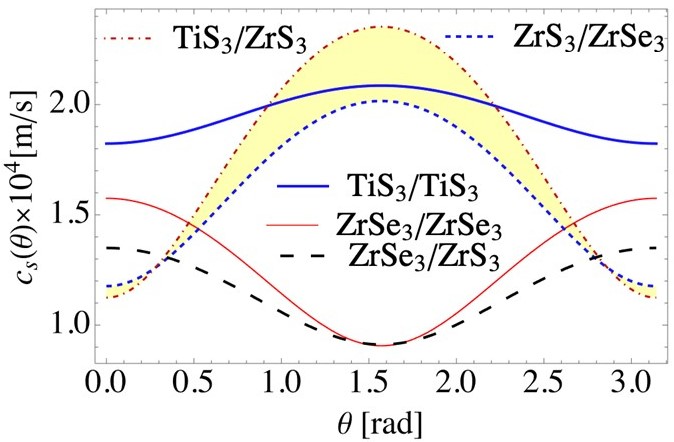}
\caption{The angle-dependent sound
velocity in different TMTC vdWHs. The interaction between the charged carriers is described by
the RK potential. The calculations were performed
for the exciton concentration $n= 4$ $\times$ $10^{15} \
\mathrm{m^{-2}}$ and the number $N_{L} = 6$ of hBN monolayers
placed between two TMTC monolayers.}
\label{fig:SoundV}
\end{figure}

The directional anisotropy of the electron and hole masses in TMTC is
reflected in the angle-dependent sound velocity. Calculations for $c_{s}(\theta )$ are performed for different TMTC vdWHs using RK potential and presented in Fig. \ref{fig:SoundV}. In Fig. \ref{fig:SoundV} the dashed-dotted and dashed curves presents $c_{s}(\theta )$ in TiS${_3}$/ZrS${_3}$ and ZrS${_3}$/ZrSe${_3}$ vdWHs, respectively, while the shaded area presents sound velocities in  ZrS${_3}$/ZrS${_3}$ and TiS${_3}$/ZrSe${_3}$. Although in TiS${_3}$/TiS${_3}$ the sound velocity has a maximum at $\theta = \pi/2$, it weakly depends on the angle due to very close values of $\mu_{x}$ and $\mu_{y}$. In contrast, $c_{s}(\theta )$ in ZrSe${_3}$/ZrSe${_3}$ and ZrSe${_3}$/ZrS${_3}$ has the same kind qualitative behavior as in phosphorene \cite{BGK2017} with minima at $\theta = \pi/2$. We also calculated the sound velocity when electron and hole interact via the Coulomb potential. It should be noted that the sound velocity is greater in the
case of Coulomb potential for the interaction between the charge carriers
than for the RK potential, because the RK potential implies the
screening effects, which make the interaction between the carriers weaker.
In the case of the sound-like spectrum of collective excitations with the angle-dependent sound
velocity $c_{s}(\theta )$ the integrals~(\ref{nxx}) and ~(\ref{nyy}) can be
evaluated analytically using \cite{Gradstein} and the results are the
following~\cite{BGK2017}
\begin{equation}
n_{n}^{xx}(T)=n_{n}^{yy}(T)=\frac{2\zeta (3)s(k_{B}T)^{3}\sqrt{M_{x}M_{y}}}{%
\pi (\hbar gn)^{2}},  \label{soundlike}
\end{equation}%
where $\zeta (z)$ is the Riemann zeta function. 
Therefore, in this case 
substituting (\ref{soundlike}) into (\ref{nNormal2}) for the concentration of the superfluid component $n_{s}(T)$ we obtain
\begin{equation}
n_{s}(T)=n-\frac{2\zeta (3)s(k_{B}T)^{3}\sqrt{M_{x}M_{y}}}{\pi (\hbar gn)^{2}}.
\label{Nor}
\end{equation}%
From Eq.~(\ref{Nor}) it follows that for the
sound-like spectrum of collective excitations, the concentrations of the
superfluid components are angle independent. Since at low momenta the sound-like energy spectrum of collective excitations of the indirect exciton in a TMTC heterostructure satisfies the Landau criterion for superfluidity, the superfluidity of IXs in this system is possible. Using Eq. (\ref{soundlike}) one can estimate the BKT phase transition temperature from the condition $n_{n}(T)=n$:
\begin{equation}
T_{BKT}=\left( \frac{\pi (\hbar g)^{2}}{2\zeta (3)s\sqrt{M_{x}M_{y}}}\right)
^{1/3}\frac{n}{k_{B}}.  \label{Tcs2}
\end{equation}%
This can be considered as an upper bound to the superfluid transition temperature. Thus, for the sound-like spectrum of collective excitations, the $T_{BKT}$ does not depend on an angle.

\bibliography{/Users/Nastia/Desktop/dissertation/bibliography/bibl_superfluidity}

\end{document}